\documentclass{llncs}

\usepackage[english]{babel}
\usepackage{float}
\usepackage{graphicx,epstopdf}
\usepackage{amsmath,amssymb,amsfonts,subfigure}
\usepackage{comment}
\usepackage{cite}

\title{A Survey of Evolving Models for Weighted Complex Networks based on their Dynamics and Evolution} 

\author{Akrati Saxena}

\institute{Department of Mathematics and Computer Science\\ Eindhoven University of Technology, Netherlands \\ 
\email{a.saxena@tue.nl}
}

\begin{document}
\maketitle

\begin{abstract}
For decades, complex networks, such as social networks, biological networks, chemical networks, technological networks, have been used to study the evolution and dynamics of different kinds of complex systems. These complex systems can be better described using weighted links as binary connections do not portray the complete information of the system. All these weighted networks evolve in a different environment by following different underlying mechanics. Researchers have worked on unraveling the evolving phenomenon of weighted networks to understand their structure and dynamics. In this chapter, we discuss the evolution of weighted networks and evolving models to generate different types of synthetic weighted networks, including undirected, directed, signed, multilayered, community, and core-periphery structured weighted networks. We further discuss various properties held by generated synthetic networks and their similarity with real-world weighted networks. 
\end{abstract}

\section{Introduction}

In real-world complex networks, each link carries a unique strength or weight \cite{latora2001efficient, banavar2000topology}. For example, in a friendship network, the weight of an edge denotes the intimacy of the relationship or frequency of the communication \cite{onnela2007analysis}. In a co-authorship network, edge weight represents the number of publications co-authored by two scientists. In an airport network, the weight of a link can represent the number of available seats, traffic flow \cite{de2005structure}, or the frequency of transport availability between two airports \cite{bagler2008analysis}. 

The origin of the concept of strength of ties dates back to 1973 when Granovetter \cite{granovetter1973strength, granovetter1983strength} introduced the idea of weak ties. He emphasized the inequality of edges in a network and broadly categorized the edges as strong ties or weak ties. Strong ties represent the relationship between people who frequently contact each other, and weak ties represent not so frequently communicated relationships. The work done by Granovetter was the first work of its kind that distinguished the edges of a network in some way. After this work, Lin et al. \cite{lin1981social} emphasized the importance of strong ties in self-growth as well as of weak ties in job achievement. In a friendship network, links can also be categorized as close friendship, acquaintance, obligation, or alliances. Researchers have verified the importance of social tie strength in real-world social networks \cite{haythornthwaite2002strong, morrison2002newcomers, brown2001granovetter, wegener1991job, ashman1998strength, yakubovich2005weak}. 

Weighted networks are characterized by a function $f(E) \rightarrow R$, where the function $f$ maps each edge to a real number, which represents the weight of that edge. Depending on the applications, weighted networks can be categorized as directed or undirected, and an edge will be denoted by an ordered or unordered pair of nodes, respectively. For example, co-authorship networks are undirected weighted networks, whereas transportation networks are directed weighted networks. In real-world networks, edges can also carry negative weights. The examples of such networks are frenemy (friend-foes) networks \cite{kunegis2009slashdot, brzozowski2008friends} or trust-distrust networks \cite{guha2004propagation, ziegler2005propagation}. The edges between two people in these networks can represent varying degrees of friendship/trust or enmity/distrust. The former kind of edges is represented by positive weights and the latter by negative weights. 

Akin to unweighted networks, weighted network also carry several properties similar to unweighted complex networks, such as scale-free degree distribution \cite{opsahl2008prominence, li2005weighted, jezewski2005scale}, small world phenomenon \cite{latora2003economic}, high local clustering coefficient \cite{opsahl2009clustering, saramaki2007generalizations}, community structure \cite{arenas2008analysis}, assortativity \cite{newman2002assortative,newman2003mixing}, and so on. Weighted networks have been used to study their evolution and analyze the dynamic phenomenon taking place on the network, such as information diffusion, opinion formation, influence propagation, and so on. However, the data collection to create weighted social networks is a computationally time taking and costly procedure, as it requires the knowledge of the strength of all relationships in a network. Therefore, researchers have studied the evolution of the topological structure of dynamically growing complex networks and proposed several modeling frameworks to generate synthetic weighted networks having properties similar to real-world weighted networks. 
The first evolving model to generate weighted networks was proposed by Yook et al. in 2001 \cite{yook2001weighted} in collaboration with Albert-László Barabasi. After this pioneering work, researchers have proposed several other models to explain various aspects of network structure based on their evolving phenomenon and corresponding environment. 

In this chapter, we will cover evolving models to generate different types of weighted networks, such as undirected and directed networks, signed networks, multilayered networks, community structured, hierarchical structured, and specialized evolving models. Most of the proposed evolving models are dynamic, i.e., the network starts with some seed nodes, and then the nodes and edges are added iteratively based on the proposed growing method. We will discuss evolving models for different kinds of networks in the coming sections. 





\section{Preliminaries}
In this section, we discuss some required definitions in the context of weighted networks. The following notations will be useful to understand given definitions. A binary network can be represented by a matrix $A$, where $a_{xy} =1$, if $x$ and $y$ are connected else $a_{xy}=0$. Similarly, a weighted network can be represented using an adjacency matrix, where $w_{xy}$ represents the weight of a link between nodes $x$ and $y$. Let $\Gamma(x)$ be the set of neighbors of node $x$ in a given network and let $k_x$ denotes the degree of node $x$. This can be modified for in-degree and out-degree in directed networks.  

\subsection{Strength of a node}

In an unweighted network, the degree of a node is defined as the total number of its neighbors. Moreover, in weighted networks, the strength of a node denotes the sum of the weight of all edges connected to that node. It is defined as,

\begin{center}
$s_{x} = \sum_{y}w_{xy}$
\end{center}

In a directed network, the in-strength of a node can be defined as the sum of the weight of incoming edges and out-strength as the sum of the weight of outgoing edges.

\subsection{Power Law Distribution}

Barabasi and Albert observed that real-world unweighted networks follow power-law degree distribution, and the probability of a node having degree $k$ is $P(k) \propto k^{-\gamma}$ \cite{barabasi1999emergence}. 

Weighted networks also follow three power-law distribution: (i) power-law degree distribution, (ii) power-law strength distribution, and (iii) power-law edge-weight distribution. 


\subsection{Preferential Attachment Model}

The preferential attachment model, also referred as BA model, is an evolutionary model proposed by Barabasi and Albert for the formation of unweighted scale free networks \cite{barabasi1999emergence}. In BA model, the probability $\prod (k_x)$ of a new node connecting to an existing node $x$ depends on the degree $k_x$ of node $x$. It is defined as,

\begin{center}
$\prod (k_x) = \frac{k_x}{\sum_{y}k_y}$
\end{center}


Analogously, in a weighted network, the nodes follow preferential attachment in accordance with the strength $s_x$ of the nodes. So,

\begin{center}
$\prod (s_x) = \frac{s_x}{\sum_{y}s_y}$.
\end{center}

The generated networks follow a `rich get richer' phenomenon that gives rise to power-law degree and strength distributions. 

\subsection{Clustering Coefficient}

Clustering coefficient of a node indicates the local density of the vicinity of a node. It is a measure of "how tightly knit your neighbors are". In unweighted networks, it can be calculated as,

\begin{center}
$CC(x) = \frac{2}{( \left | x \right |) *  (\left | x \right | - 1)}\sum a_{xy}a_{yz}a_{xz}$
\end{center}
Barrat et al. \cite{barrat2004architecture} defined clustering coefficient for weighted networks as,

\begin{center}
$CC(x) = \frac{1}{s_{x} * (\left | x \right | - 1)}\sum_{y,z}\frac{w_{xy}+w_{xz}}{2} a_{xy}a_{yz}a_{xz}$
\end{center}
Zhang et al. do the analytical studies of this and shows its dependency on node degree and strength \cite{zhang2009analytic}. 
Lopez et al. \cite{lopez2004applying} considered the total weight of relationships while defining the clustering coefficient. According to the proposed formula,

\begin{center}
$CC(x) = \frac{\sum_{y \neq z \epsilon \Gamma(x)} w_{yz}}{\left | x \right | * (\left | x \right | - 1)}$
\end{center}
There are some other generalizations of clustering coefficient introduced by Onnela et al. \cite{onnela2005intensity} , Zhang et al. \cite{zhang2005general} , Schank et al \cite{schank2004approximating} , Holme et al. \cite{holme2007korean}, Kalna et al. \cite{kalna2006clustering} and Opsahl et al. \cite{opsahl2009clustering}.

\section{Undirected and Directed Weighted Networks}

The first work to propose evolving models for generating synthetic weighted networks dates back to 2001, when Yook et al. \cite{yook2001weighted} proposed two evolving models, (i) Weighted Scale Free (WSF) model and (ii) Weighted Exponential (WE) model. In both models, the network starts with a seed network having $n_0$ nodes. In the WSF model, every new incoming node is connected with $m$ nodes using the degree preferential attachment rule. Each new node $x$ has fixed weight $w_x=1$, and this weight is distributed among its all $m$ edges. The weight distribution across each edge $(x,y)$ is proportional to the degree of other endpoint as $w_{xy} \propto k_y$. 

\begin{center}
$w_{xy} = \frac{k_y}{\sum_{y'}k_y'}$
\end{center}
where $y'$ belongs to $m$ chosen neighbors of the new node $x$.

In the WE model, a new node makes $m$ connections with the existing nodes with equal probability, and the weight assignment for the edges is the same as in the WSF model. The authors show that the generated networks using both the models follow power-law degree, strength, and edge-weight distribution. After this work, several evolving models have been proposed, and we have categorized them as discussed in the following subsections.

\subsection{Power Law Minimal Model}

In real-world weighted networks, degree, strength, and edge-weights follow a power-law distribution. 
In this section, we will discuss evolving models based on power-law properties that can be designed with minimal effort, therefore referred to as \textit{minimal models}.

In 2005, Antal et al. \cite{antal2005weight} proposed a model in which the network starts with some seed nodes, and each new incoming node $x$ makes an edge with an existing node $y$. The edge weight $w_{xy}$ is chosen uniformly from the given edge weight distribution $\rho(w)$ and remains fixed once assigned. They further introduced two generalizations of this model, (i) each new node makes $m$ connections in the network, and (ii) after adding a new node to the network, some links are also added among already existing nodes. The probability of adding a link $(x,y)$ is directly proportional to the strength of its endpoints $P(x,y) \propto s_x \cdot s_y$. The authors showed that in the generated networks, the strength distribution has a  tail ,i.e., independent of the edge-weight distribution. 

Bianconi \cite{bianconi2005emergence} studied the evolution of weighted networks and divided these networks into two main classes based on the ratio of the rate of strengthening of edge weights and the rate of making new connections. In traffic networks, new links are formed more frequently with new nodes than increasing the strength of the already existing links. This type of networks is called class $1$ networks, and there is a linear relationship between the strength and degree of a node, $s_x \propto k_x$. While in co-authorship networks, two existing collaborators publish more papers together than with a new collaborator. This type of networks is called class $2$ networks and follow non-linear relation as $s_x \propto k_x^\theta$ where $\theta >1$. This model uses two steps, at every timestamp, (i) a new node is added and make $m$ connections using degree preferential attachment, and (ii) choose $m'$ already existing edges preferentially and increase their weights by $w_0$. To choose $m'$ edges, the model first chooses a node using degree preferential and then chooses one of its edges with probability proportional to the edge weight. We can set different values of $m$ and $m'$ to generate class $1$ and class $2$ networks. 

Mukherjee et al. \cite{mukherjee2006weighted1} studied the weighted model based on the self-organizing link weights dynamics, and the following two steps govern the network growth.
\begin{enumerate}
\item The network starts with a seed graph. At every timestamp, one edge is selected randomly, and a node is connected to one of its endpoints with equal probability.
\item After adding a new node, edge weights are updated using the following two steps,
\begin{enumerate}
\item $s_x=\sum_yw_{xy}$, and
\item $w_{xy}=(s_xs_y)^\alpha$, where, $\alpha$ is a parameter and its value can vary for different applications.
\end{enumerate}
\end{enumerate}
These networks follow all three power-law distributions observed in weighted networks. \cite{chen2007origin} discussed a generalized preferential attachment probability function to generate four power-law distributions, including degree, strength, link weight, and subgraph degree.

\subsection{Fitness Model}

In the above-discussed models, the number of links a node attracts is proportional to its degree or strength. However, in many real-world networks, this phenomenon is not enough to explain the network evolution and its dynamics. 
In 2001, Bianconi et al. \cite{bianconi2001competition} introduced the concept of fitness for unweighted networks where each node has a fitness value $\eta$ and more fit nodes attract more links; this phenomenon is also called ``fitter gets richer".

Wang et al. \cite{wang2004weighted} proposed a fitness based model called Weighted Competition Scale Free (WCSF) model that is quite similar to the BA model. The network starts with a seed graph, and at every timestamp, 
\begin{enumerate}
    \item With probability $p$, a new node $x$ is added to the network with $m$ edges connected preferentially. This new node has fitness value $\eta_x$, that is chosen uniformly from a fixed power-law fitness distribution $\rho(\eta)$ in the range $[\eta_{min}, \eta_{max}]$ and $\eta_{min}>0$. The probability of selecting an existing node $y$ to make a new connection considers the fitness of the node. It is defined as,
\begin{center}
$\prod_y = \frac{\eta_yk_y}{\sum_{z}\eta_zk_z}$
\end{center}
\item With probability $(1-p)$, the network will be self-grown, as no new node will be added, and $m$ edges will be added among the existing nodes. To add a new edge, two end nodes are selected using the fitness preferential attachment rule. If they are not connected, connect them with edge weight $w_0=1$, otherwise increase the edge weight by $w_0=1$. 
\end{enumerate}

Thus, in the fitness model, the nodes having high fitness attract more new and self-growing links. 

Zheng et al. \cite{zheng2003weighted} presented a stochastic weighted model where each node has a fitness value $\eta$ uniformly distributed over range $[0,1]$. In this model, with probability $p$, the edge-weights are assigned using the WSF model, and with probability $(1-p)$, edge-weights are assigned using the fitness parameter. The link weight is directly proportional to the fitness parameter of its end point  $w_{xy} \propto \eta_y$, so,
\begin{center}
$w_{xy} = \frac{\eta_y}{\sum_{y'}\eta_{y'}}$
\end{center}

New connections can also consider some more intrinsic properties of the nodes. \cite{geng2007weighted} proposed a model based on the intrinsic strength of the nodes.  

\subsection{Stochastic Model}

In \cite{riccaboni2010structure}, the authors proposed a simple stochastic model by incorporating weight dynamics in the BA model. The authors assume that the link weights are evolved as the geometric Brownian motion, also called Gilbrat's law of proportionate effects. To understand weight dynamics, they used the theoretical framework of the scaling distribution of fluctuations proposed by Stanley et al. \cite{yamasaki2006preferential, fu2005growth}. The model starts with a seed network having $n_0$ nodes, each with a self-loop. At every timestamp $t$, a node $x$ is chosen with probability $k_x(t-1)/2t$ to make a new connection. With probability $\alpha$, a new node is added to the network, and node $x$ is connected to this. With probability $(1-\alpha)$, node $x$ is connected to already existing node $y$, where node $y$ is chosen with probability $k_y(t-1)/(2t-k_x(t-1))$ if $x \neq y$, otherwise probability is $0$. The edge-weight is assigned randomly from a fixed power-law distribution. After adding a new edge, the weight of each edge is decreased or increased by a random factor $\chi_{xy}(t)$, so edge weight at time $(t+1)$ is $w_{xy}(t+1) = w_{xy}(t)\chi_{xy}(t)$. Thus, the weight of edges grows randomly with time, and it is similar to geometric Brownian motion. The model is verified using NBER-United Nations Trade Data \cite{feenstra2005world}. The network evolves from random to exponential to scale-free graphs as the value of $\alpha$ increases. 


\subsection{Incremental Weight-Distribution Model}

In traffic networks, the addition of a new edge to an existing node increases the traffic on that node. This increased traffic is distributed among the neighbors of the node; such a distribution is called the local rearrangement of the weights.
In 2004, Barrat et al. \cite{ barrat2004weighted, barrat2004modeling, barthelemy2005characterization} proposed a weighted traffic evolving model based on this phenomenon, also called BBV (Barrat, Barthelemy, and Vespignani) model. This model has the following two steps.
\begin{enumerate}
\item The model starts with completely connected $n_0$ nodes, and each edge has weight $w_0$.
\item  At each timestamp, a new node $x$ is added and connects with $m$ nodes having weight $w_0$ using strength preferential attachment rule. Each new connection adds some extra traffic $\delta$ to its endpoint node, so, 
\begin{center}
$s_{y} = s_{y} + w_{xy} + \delta $
\end{center}
This extra created traffic is distributed to all the neighbors of the node proportional to their edge weights. Now, the new weight of an edge $(y,z)$ is computed as,

\begin{center}
$w_{yz} = w_{yz}+ \delta\tfrac{w_{yz}}{s_y}$
\end{center}

\end{enumerate}
If $\delta = w_0 =1$, it perfectly depicts the traffic model, where, all incoming traffic is mostly distributed on outgoing links. If $\delta<1$, it is similar to collaboration network, where a new collaboration does not affect old collaborations of the node. $\delta>1$ signifies the situation where incoming edge bursts more traffic on other neighbors. The authors further studied the evolution of the strength and degree of the nodes as well as of the edge-weight. The strength of a node will be changed if the link is directly connected to this node or one of its neighbors. So, the average value of the strength of a node $x$ at time $t$ $(s_x(t))$ can be calculated using evolution equation for $s_x$,
\begin{center}
$\frac{\mathrm{d} s_x}{\mathrm{d} t} = m\frac{s_x}{\sum_{y}s_y}(1+\delta) + \sum_{y\epsilon \Gamma (x)}m\frac{s_y}{\sum_{z}s_z}\delta\frac{w_{xy}}{s_y}$
\end{center}

Similarly, the degree of a node $x$ is evolved as,

\begin{center}
$\frac{\mathrm{d} k_x}{\mathrm{d} t} = m\frac{s_x(t)}{\sum_{y}s_y(t)}$
\end{center}

Similar to the strength evolution, the weight of an edge changes when a node is directly connected to it or one of its neighbors. The rate of change of edge weight $w_{xy}$ can be computed as:
\begin{center}
$\frac{\mathrm{d} w_{xy}}{\mathrm{d} t} = m\frac{s_x}{\sum_{y}s_y}\delta\frac{w_{xy}}{s_x} + m\frac{s_y}{\sum_{y}s_y}\delta\frac{w_{xy}}{s_y}$
\end{center}
 
In the networks generated using BBV Model, the degrees, strengths, and edge weights follow a power-law distribution.
 
Hu et al. \cite{hu2005generalized} extended the BBV model \cite{barrat2004weighted} and proposed two separate models for traffic and friendship weighted networks. In the traffic driven model, they take a constant traffic increasing rate $W$ for a network. This new traffic is distributed among all the edges proportional to their strength. The rest of the mechanism of this model is similar to that of the BBV model. At every timestamp, the extra added traffic on each node is computed as,

\begin{center}
$\Delta W_x = W\frac{s_x}{\sum_{y}s_y}$
\end{center}

So, the time evolution equation of edge weight is given as,

\begin{center}
$\frac{\mathrm{d} w_{xy}}{\mathrm{d} t} =\Delta W_x\frac{w_{xy}}{s_x} + \Delta W_y\frac{w_{xy}}{s_y}$
\end{center}
 
Time evolution equation of strength is given as,
 \begin{center}
$\frac{\mathrm{d} s_x}{\mathrm{d} t} = \sum_{y\epsilon \Gamma (x)} \frac{\mathrm{d} w_{xy}}{\mathrm{d} t}   +  m\frac{s_x}{\sum_{z}s_z}$
\end{center}

Time evolution equation of degree is given as,

\begin{center}
$\frac{\mathrm{d} k_x}{\mathrm{d} t} = m\frac{s_x}{\sum_{y}s_y}$
\end{center}

In the second model, they explained the dynamics of friendship formation using ``friends of friends" phenomenon. Network formation starts with a seed graph, the same as in BBV model. At every timestamp: 
\begin{enumerate}
\item A new node $x$ is connected to a node $y$ using strength preferential attachment. 
\item Node $x$ is connected to $m$ neighbors of node $y$ with the weight preferential probability, given as,

\begin{center}
$P(x,z) = \frac{w_{yz}}{s_y}$
\end{center}
where $z$  $\epsilon$  $\Gamma(y)$.
\end{enumerate}

The first link is called the primary link, and the remaining $m$ links are called secondary links. To control the effect of newly added links, they do the local rearrangements of weights the same as BBV model. This model perfectly illustrates the real-world social networks or collaboration networks. If a new person $A$ forms a link with a person $B$, then there is a very high probability that $A$ will make links with $B's$ friends.

Goh et al. \cite{goh2004traffic} proposed a very specific model for traffic networks. In traffic networks, the weight of a link can be calculated by measuring the load of the traffic on that edge \cite{goh2001universal} or betweenness centrality of that edge \cite{saxena2020centrality}. The edge-weight also depends on the degree of its endpoints as $w_{xy} \sim (k_xk_y)^\theta$. The load of a node is calculated in a similar manner as the strength of a node; it is the sum of the load of all edges connecting to that node. The load of a node is directly proportional to the strength of the node $l_x \propto s_x$. The model starts with the seed graph the same as mentioned in the BBV model. At every timestamp, a new node is added with $m$ edges using sub-preferential attachment probability as,

\begin{center}
$P(x) \sim s_x^\alpha \sim l_x^\alpha$
\end{center}
where, $\alpha = 1/\eta$ and $l_x \sim k_x^\eta$. The load of each node is recalculated after every step. The networks generated using this model follow strength and degree power laws. This model gives a new approach to understand the evolution of weighted networks with the increasing load in the system. Each new node and link can affect the shortest pathway in the networks. So, link weights are updated regularly. Some other networks, such as a friendship network or co-authorship network, where the edge weight depends on the harmony of their relationship, can be best explained using BBV model.


Some other works that studied the slight variations of BBV model include \cite{junfen2011weighted, fu2004weighted, geng2009evolving, liu2007evolving, su2009flexible, chrysafis2009weighted, fang2007synthetical, hui2006self, jian2006evolution, she2009model}; these models proposed different weight distribution function using some constraints. \cite{gui2009mixed} discussed a model based on weight driven and inner selection mechanism. Lu et al. \cite{lu2006topological} studied the combination of both deterministic and random preferential attachment and proposed a hybrid preferential attachment (HP-BBV) model.

\subsection{Node-Deactivation Model}

Wu et al. \cite{wu2005generating} presented a model inspired by the concept of degree dependent node deactivation model \cite{klemm2002highly} and the strength based traffic distribution \cite{pandya2004note}. This model considers that each node can have either of the two states: active or inactive. Pandya proposed the strength based traffic distribution idea where increased traffic moves towards the bigger airports that can handle it \cite{pandya2004note}, and thus the traffic on high strength node gets increased. In \cite{wu2005generating} model, at every time step, $m$ nodes are in active state and a new node is connected to these $m$ active vertices. The increased traffic on connected nodes is distributed as, 
\begin{center}
$w_{xy}=w_{xy}+ \delta \frac{s^{in}_y}{\sum_{z \epsilon \Gamma(x)}s^{in}_z}$
\end{center}
where $s^{in}_x$ is the in-strength of node $x$. The newly added node will always be in the active state in the next time step. So, there are total $(m+1)$ active nodes and one active node will be converted to inactive node with the probability given as,
\begin{center}
$\prod(s^{in}_x) \propto \frac{\gamma -1 }{a+ s^{in}_x}$
\end{center}
where $a>0$ and $\gamma -1 = \frac{1}{\sum_{y \epsilon M}1/(a+s^{in}_y)}$, $M$ is the set of all active nodes in the last time step. It shows that a high strength node has low probability to be deactivated in the next time step. The constant parameter $a$ is used as a bias factor and affects the power law exponent by varying the probabilities of node deactivation. 

Tian et al. \cite{tian2007effect} studied the deactivation mechanism using the total strength of the nodes. They proposed a different probability function to deactivate a node, given as,
\begin{center}
$\prod(s_x) = \frac{\alpha}{s_x}$
\end{center}
where, $\alpha = \frac{1}{\sum_{y \epsilon M}1/s_y}$.
They also showed that the average clustering coefficient decreases as the number of active nodes increases, so $C \propto 1/M$. Due to the deactivation phenomenon, these networks possess disassortative behavior and hierarchical organization. These models can help us in a better understanding of disassortative networks. These networks also follow all weighted network power-law distributions. 

Ma et al. \cite{ma2013weighted} explained the social evolution model using the self-organized birth and death process of human life. The network size increases gradually if the birth rate exceeds the death rate as in natural evolution. Model is based on three natural characteristics of natural evolution, (i) reproduction, (ii) cooperation, and (iii) competition. At each time step, one of the following steps is processed.
\begin{enumerate}
\item With probability $p$, an offspring is created and connected to one randomly chosen parent. Offspring inherits $m_1$ links preferentially from its parent.
\item With probability $q$, $m_2$ links of chosen parent is deleted. 
\item With probability $r$, $m_3$ new links are added to the network preferentially.
\item With probability $s$, a chosen node competes with its neighbor.
\end{enumerate}
where $p+q+r+s=1$. This is a basic model to explain the birth-death process and can be extended further to explain more real life events.

\subsection{Non-Linear Growing Model}

In some real-world networks, as new nodes keep coming, the rate of adding new connections by the newly added node increases exponentially. The reason is that the network size increases with time, and a new coming node has more options to make connections. This type of networks is called \textit{exponentially growing networks} as the total number of edges grows exponentially with time. In these networks, when a new node joins at timestamp $t$, the number of connections made by this new node follows the power-law function. So, the number of edges added by this new node is $t^\theta$, where $\theta$ is called acceleration parameter and $(0 \leq \theta \leq 1)$. Zhang et al. \cite{zhang2009effects} proposed a model to capture the evolution of such growing networks. In the given model, at time $t$, the network has $(t + n_0)$ nodes ($n_0$ is the number of nodes in the seed graph) and $\int t^\theta \mathrm{d}t =t^{1+\theta}/(1+\theta)$ edges. The weight rearrangement dynamics is the same as in the BBV model. The authors showed that the value of accelerating exponent is 0.56, 0.18, and 0.12 for arXiv citation graph, autonomous Internet graph, and the email network, respectively. By setting different values of $\theta$, one can get scale-free to exponential graphs, and the degree, strength, and edge weight distribution changes from a small exponent to a large exponent. The clustering coefficient also depends on the value of $\theta$, and as $\theta$ increases, the clustering coefficient also increases. If $\theta =1$, the model converts to the basic weighted scale-free model.

\cite{wang2008extensive} studied the non-linear growing network and propose two models to reproduce these networks. Rui et al. \cite{rui2012nonlinear} also studied the non-linear growth of edge weight in the local world environment. Each new edge, added at timestamp $t$, is assigned weight $w_0=t^\theta$, where $0 \leq \theta <1$. This model is a good generalization of the BBV model to capture the peculiarities of both kinds of networks. Eom et al. \cite{eom2008evolution} proposed a linear and non-linear growth model for online-bulletin board and movie-actor network. Dai et al. \cite{dai2014dynamic} studied the phenomenon, where at each time step, edge weights are increased or decreased non-linearly using a given probability function. \cite{sen2007weighted} proposed a model where each new node makes edges proportional to the size of the network $m \propto N$. Another variation of these models is \cite{mu2014weighted}.

\subsection{Incremental Self-Growing Model}


Real-world networks self-grow, where the already existing nodes keep making new connections. For example, in a co-authorship network, new researchers will join and make connections with the existing nodes, and in the meantime, already existing researchers will also build new collaborations. In this section, we discuss models that capture both topological growth as well as self-growth.

In 2005, Wang et al. \cite{wang2005mutual} proposed the first self-growing model for weighted networks. The network starts with $n_0$ fully connected nodes, and each edge is having weight $w_0$. the network grows by following two given mechanisms, at every timestamp,
\begin{enumerate}
\item \textbf{Topological Growth:} A new node joins the network and connects with $m$ nodes using strength preferential attachment.
\item \textbf{Mutual Selection Growth:} Each existing node $x$ selects $m'$ other nodes with the given probability function, as:

\begin{center}
$P(x,y) = \frac{s_y}{\sum_{z}s_z -s_x}$
\end{center}

If the chosen node $y$ is already connected to $x$, their edge weight is increased by $w_0$. If they are not connected, an internal link $(x,y)$ is formed with edge weight $w_0$.
\end{enumerate}

The evolution equation of edge weight can be written as,
\begin{center}
$\frac{\mathrm{d} w_{xy}}{\mathrm{d} t} = m\frac{s_x}{\sum_{z}s_z -s_x} \times m\frac{s_y}{\sum_{z}s_z -s_y}$.
\end{center}

The strength of a node $x$ can be changed if the new node is connected to it or this node is chosen for mutual selection growth. Th strength evolution equation is defined as,
\begin{center}
$\frac{\mathrm{d} s_x}{\mathrm{d} t} = \sum_{y}\frac{\mathrm{d} w_{xy}}{\mathrm{d} t} +n\times \frac{s_x}{\sum_{z}s_z}$
\end{center}

This equation's solution shows that the total strength of nodes is uniformly increased with the network size. The authors showed that the model depicts the disassortative nature of real-world networks.

Xie et al. \cite{xie2007modeling} extended the work in \cite{wang2005mutual} and proposed a strength dynamic model where all edges are strengthen up continuously and new connections are added among existing nodes. In this model, the topological growth works same as in the \cite{wang2005mutual}. However, they proposed a different approach for strength dynamics. At each time step, all edges update their weight using the strength coupling mechanism, given as, 
\begin{center}
$w_{xy} = \left \{\begin{matrix}
w_{xy}+1 &  with & probability & Wp_{xy}\\
w_{xy} &  with & probability & 1- Wp_{xy}
\end{matrix}\right.$
\end{center}
where,
\begin{center}
$p_{xy}=\frac{s_xs_y}{\sum_{m<n}s_ms_n}$
\end{center}
$W$ shows the increasing rate of edge weight. If the value of $Wp_{xy}$ is greater than $1$, then it is assumed to be $1$. A high value of $W$ results in a higher clustering coefficient. Mu et al. \cite{mu2010novel} proposed a different probability function for the same model as,
\begin{center}
$p_{xy}=\frac{2w_{xy}}{\sum_{z}s_z}$
\end{center}
After updating the weight of all edges, newly added edges create an extra flow to the connected nodes. \cite{hui2007weighted} propose a local weight rearrangement mechanics to handle this extra traffic.


Tanaka et al. \cite{tanaka2007scale} presented a different self-growing strategy to make or strengthen the links among already existing nodes. In this model, at every timestamp $t$, $ct$ ($c$ is a constant) pairs of nodes are selected for the self-growth using the probability proportional to their strength, as,

\begin{center}
$P(x,y) = \frac{s_xs_y}{(\sum_{z}s_z)^2}$
\end{center}

In this model, the total number of nodes is $n \approx t$, so the strength of a node will be increased by $\Delta s_x \approx 2c$ when a new node is added. They also showed that for a coauthorship network $c = 1.5 \times 10^{-4}$. The lower value of $c$ shows that the rate of joining of a new author is frequent as compared to the already existing scientists collaborating. The communication networks, such as email or friendship networks, where new nodes join very rarely, have a high value of $c$.

They further studied the evolution of weighted networks and proposed an approach to make scale-free networks with variable power-law exponent \cite{tanaka2008weighted}, where different values of the given parameters result in different power-law exponents. They defined the strength driven preferential attachment probability as, 
\begin{center}
$P(x) = \frac{(s_x+\sigma)}{\sum_{z}(s_z + \sigma)}$
\end{center}

where, $\sigma$ is a constant. In this model, the network starts with a single node. At each time step, a new node is added with $m$ edges having weight $1$ using the strength driven preferential attachment probability as given above. At each time step, $ct^{\eta}$ pairs of nodes are chosen for self-growth with the probability defined as,
\begin{center}
$P(x,y) = \frac{(s_x + \sigma)(s_y+\sigma)}{(\sum_{z}s_z + \sigma)^2}$
\end{center}
After choosing a pair of nodes, links are added similar to other models. As a new node has maximum $m$ edges, the strength of the new node is $m$. The value of $\sigma$ is defined as $\sigma > -m$ so that a new node can also attract more edges in the next time step. The value of parameter $c$ decides the density of the network, and this model shows exponential growth with time.

Real-world weighted networks can be assortative or disassortative. Leung et al. proposes a model to generate both kind of networks by setting value of given parameters \cite{leung2007weighted}. In this model, at timestamp $t$, a new node is added to the network and makes $pt^\theta$ links preferentially, where $p>0$ and $0< \theta < 1$. They modified the self growth procedure to get assortativity in the network. At every timestamp, a link $(x,y)$ is chosen preferentially and its weight is updated as, 
\begin{center}
$w_{xy}=w_{xy}+sgn(q)$
\end{center}
where,
\begin{center}
$sgn(q)=\left\{\begin{matrix} 1 & if & q>0\\ -1 & if & q<0\\ 0 & if & q=0 \end{matrix}\right.$
\end{center}
When a link has weight $0$, it is removed from the network. For each timestamp $t$, self growth process is repeated $|q|t^\theta$ times. As the value of $q$ increases, network topological behavior changes from assortative to disassortative.

\subsection{Triad-Formation Model} 

All real-world networks have a self-growing phenomenon, and Wang et al. \cite{wang2005mutual} captured this growing phenomenon very well. Hao et al. \cite{hao2009general} studied the self-growing phenomenon of real-world networks in detail and proposed a model based on their findings. When a new node joins the network, it is connected to some existing nodes. With time, the already existing nodes also make some connections, but in the real world, these new connections are mostly based on \textit{friends of friends} phenomenon. In real-world networks, people prefer to make connections with the neighbors of its neighbors, also called \textit{triad formation} (TF).  
The proposed model works as follows: 
\begin{enumerate}
\item The network starts with a seed graph as in the other models.
\item With probability $p$, a new node is added to the network with $m$ edges using the preferential attachment. After adding $m$ edges, the local rearrangement of weights is done as per BBV model.
\item With probability $(1-p)$, network is self-grown using $m$ edges. Self-growing links can be of two types: (i) Triad Formation (TF) link or (ii) Random link. With probability $\varphi$, a TF link is added using TF Rule, and with probability $(1-\varphi)$ a random link is added. To make a triad link, first select an edge $(x,y)$ randomly. Then select one other neighbor $z$ of $y$ using the following preferential attachment rule to make the triad. The probability to choose other node is defined as,
\begin{center}
$\prod(z) = \frac{w_{yz}}{(s_y - w_{xy})}$
\end{center}
If there is no link between $x$ and $z$, make a link having weight $w_0$; otherwise, increase its weight with some constant $\sigma$.
\end{enumerate}
This modeling approach can be used to design networks with varying clustering coefficient by tuning probabilities of making PA, TF, and random links. The higher the probability of making triad links, the higher is the clustering coefficient. Zhang et al. \cite{zhang2010weighted} studied this process with PA and TF links. These networks also possess high clustering coefficient, and average clustering coefficient gives power-law distribution with the degree of nodes, $C(k) \propto k^{-\gamma}$.
Some more studies have been performed on the triad formation based model, including \cite{jing2009general, wang2012effect, yang2014local, li2011weighted, dan2012synchronizability}.

\subsection{Traffic-Flow Driven Model}

In traffic networks, edge weights can be calculated using traffic flow or betweenness centrality measure. In this section, we discuss a few models related to these approaches.

Hu et al. \cite{hu2007weighted, hu2008weighted} studied the pattern of passenger behavior on the transportation network. In an airline network, they found that passengers either choose a direct route to the destination or by one hop. This middle hop mostly belongs to the hubs of the network. They propose a model where a new node is directly connected to the destination with probability $p$ or by passing through a third node with probability $(1-p)$. This middle transfer node $y$ on the path $x \rightarrow z$ is chosen using edge weight preferential attachment, defined as, 
\begin{center}
$\prod(x,y) = \frac{w_{zy}}{s_z}$
\end{center}
This human behavior gives birth to disassortativeness in transportation networks. These networks have high clustering coefficients when $p$ is small; this shows that transform behavior is responsible for cluster formation and hierarchical organization.

Dai et al. \cite{dai2013weighted} proposed a model for the traffic flow of the transportation network. This model is based on deterministic and random node attachment and shows scale-free properties. \cite{zheng2008weighted} studied the flow of traffic with network topology. In this model, nodes represent the traffic flow states, and edge-weight represents the transported traffic flow between two different traffic flow states. Each node contains a set of cells as each road can be divided into some cells. The link-weight depends on both the state and velocity of the cells. For more detail, \cite{zheng2008weighted} can be referred. \cite{jian2007strength, hu2009cost, bing2006weighted} also studied traffic driven models with distinct parameters effect, such as congestion effect.
 
\subsection{Local World Model}

In the above-discussed models, a newly added node makes connections globally or using the triad-formation rule; however, in real-life, each node belongs to a group or community referred to as its local world. 
Li et al. \cite{li2003local} introduced the concept of local world in unweighted networks. These local world evolving networks shows a transition between exponential and scale-free degree distribution. They further proposed a weighted network model based on the concept of local world \cite{pan2006generalized}. In this model, a network starts with the seed graph. For every new coming node, we follow the next given steps:
\begin{enumerate}
\item To determine the local world of a new coming node, we choose $M$ nodes randomly from the network, and this group of nodes is called the local world of the new node.
\item the new node is connected to $m$ nodes in its local world using the preferential attachment. the probability of connecting new node $x$ to node $y$ is given as,
\begin{center}
$P(x,y) = P'(y \epsilon M)\frac{s_y}{\sum_{z \epsilon M}s_z}$
\end{center}
where, $P'$ is the probability that node $y$ belongs to the randomly chosen local world and $P'(y \epsilon M) = M/n(t)$, $n(t)$ shows the total number of nodes in the graph at time $t$.
\item In the last step, local rearrangement of the edge weights are done to handle the extra created traffic by new edges. Weights can be rearranged in two ways: (i) using the BBV model rearrangement, (ii) using Hu et al. model \cite{hu2005generalized} except here the constant increased weight is distributed only among the chosen local world nodes proportional to their strength. So, the increased weight of a node $y$ $(y \epsilon M)$ is,
\begin{center}
$\Delta W_y = W\frac{s_y}{\sum_{z \epsilon M}s_z}$
\end{center}
This extra introduced traffic on each node $\Delta W_y$ is preferentially distributed to all the connecting edges.
\end{enumerate}
In this model, if the new node is connected to all nodes of the local world $(m = M)$, degrees, strengths, and edge weights show exponentially decaying distribution. If the local world is chosen as the whole network, this model gives the same result as BBV model. When we increase the size of $M$, the model shows a transition from assortative networks to disassortative networks.

Sun et al. \cite{sun2007unweighted} studied this phenomenon in more detail and presented a weighted model using local information that exhibits the transition from unweighted to weighted networks. In their proposal, at every timestamp, when we determine the local world, one of the following two procedures can happen. With probability $p$, the network has increment growth by adding a new node with $m$ edges in this local world. With probability $(1-p)$, self-growth happens by adding $m$ edges in the determined local world. So, in $t$ timestamps, $pt$ nodes and $mt$ edges are added to the network. This model depicts the real-world phenomenon where high strength nodes keep strengthening their relation, and at the same time, they also make connections with new nodes. The time evolution equation of edge weight is given as,
\begin{center}
$\frac{\mathrm{d} w_{xy}}{\mathrm{d} t} = pm\prod_{local}(n \rightarrow x)\frac{1}{N} + (1-p)\prod_{local}(n \rightarrow x)\prod_{local}(n \rightarrow y)$
\end{center}
where $N$ is the total number of nodes, $\prod_{local}(n \rightarrow x)$ is the probability that a new node is connected to node $x$ when $x \epsilon M$. They also studied synchronization dynamics on the generated networks \cite{barahona2002synchronization}. They found that the network synchronization can be increased by reducing the heterogeneity of the edge weights.

To improve the clustering coefficient of the above suggested model, Zhang et al. \cite{zhang2007local} included the concept of triad formation. They suggested that when a new node is added to its local world preferentially, it will also make some triad links (be friend with friends of friends) \cite{holme2002growing}. This phenomenon happens in real life and helps in improving the clustering coefficient and average path length (small world) of the network. The model starts with the seed graph of $n_0$ nodes and $m_0$ edges. At every time step, a new node is added, and it makes $m$ preferential edges in its local world. After adding each PA link, with probability $p$, the new node also makes a TF link with any random neighbor of the last chosen node using PA law. By changing the value of $p$, we can get networks of varying clustering coefficient. After $t$ time step, the network has $(n_0+t)$ nodes and $(1+p)mt+m_0$ edges. This model gives logarithmic average path length with the total number of nodes.

In 2007, Li et al. \cite{li2007evolving} proposed a model based on the similar concept that nearest neighbors or friends of friends have a high probability of being your friend. The model is inspired by scientific collaboration networks. To use this information, they use a path-based preferential attachment. The probability of connection increases as the shortest path length decreases. Edge weight is defined as a function of the connecting time of that edge. If node $x$ and $y$ are connected at time $T_{xy}$ then edge weight $w_{xy}$ is defined as: 
\begin{center}
$w_{xy} = f(T_{xy})$
\end{center}
This function can be of different types but they used linear function $w_{xy} = \alpha T_{xy}$ for the simulation. The model starts with the fully connected seed nodes and all edges having weight $w_{xy} = f(1)$. The initial time is set to $1$, and at every timestamp, steps from 1 to 4 are followed as explained next:
\begin{enumerate}
\item A new node is added to the network, and $l$ old nodes are chosen randomly.
\item Every selected node ($x$) is assumed to be ready to make a connection. The probability of making an edge from node $x$ to node $y$ is defined as:
 \begin{center}
$P(x,y)=(1-p)\frac{k_y}{\sum_{z}k_z}+(p-\delta)\frac{s_y}{\sum_{z}s_z}+\delta\frac{l_{xy}}{\sum_{z\epsilon\partial_x^{1,2}}l_{xz}}$
\end{center}
where, $l_{xy}$ is the similarity distance between node $x$ and $y$, and $\partial_x^{d}$ is the set of $d$ distance neighbors of node $x$. In the given probability function, they only used $1$ and $2$ distance neighbors because they have high probability to encounter and be friends.
\item After selecting a node $y'$ using above probability function, $x$ and $y'$ are connected and their connecting time is increased as: 
  \begin{center}
$T_{xy'}(t+1)=T_{xy'}(t)+1$
\end{center}
\item The edge weight is updated as:
\begin{center}
$w_{xy'}(t+1)=f(T_{xy'}(t+1))$
\end{center}
\end{enumerate}
The networks generated using this model follows strength and degree power law distribution. These networks also have a very high clustering coefficient because of using similarity distance. This model can be generalized to get directed weighted networks. \cite{cao2006neighbourhood} studied a similar local world model by considering the neighborhood of depth $d$ as the local world of a new coming node. \cite{sen2009new} extended local-world model and proposed a model for the scenario where all nodes do not have their local world information. \cite{wen2009dynamics} introduced the concept of weak ties between different local worlds.

\subsection{Mutual Attraction Model}

In real-world social networks, the friendship between two people depends on their mutual affinity, intimacy, attachment, and understanding. None of these parameters is considered in any above-discussed model. In friendship networks, each person has some attractiveness to grab more friends. A new person with a highly attractive nature can have more friends than an old less attractive person. Attractiveness can be defined as the nature of the person, talkativeness, understanding, confidence, loyalty, etc. For example, in the co-authorship network, the attractiveness of an author can not be described by only the number of publications. It will depend on other parameters, such as enthusiasm, openness to work with others, research environment, and discipline, etc. Dorogovtsev et al. \cite{dorogovtsev2000structure} proposed a model based on the initial attractiveness of websites.

Wang et al. \cite{wang2006mutual} proposed a model based on the attractiveness and mutual attraction of two people. The model starts with $n_0$ isolated nodes; each having initial attractiveness $A$, $(A>0)$. At each timestamp, a new isolated node $x$ is added to the network. Then every node in the network chooses $m$ other nodes as the perspective available options to make connections. Node $x$ chooses node $y$ with the probability given as,
\begin{center}
$\prod(x,y)=\frac{s_y+A}{\sum_{z \neq x}s_z+A}$
\end{center}
However, the selection does not guarantee a connection. Now, if two nodes have chosen each other mutually, their edge weight is increased by $1$ if they are not connected; otherwise, an edge having weight $1$ is formed. The higher the value of $A$, the higher the chances for new nodes to get more connections. Parameter $m$ controls the density of the network. As the value of $m$ increases, each node selects more options, and therefore, hubs have a high probability of getting more connections. A higher value of $m$ leads to disassortative networks. This model works on common interests and mutual acknowledgments. The edge weight is updated if both nodes select each other, so the time evolution equation of edge weight can be written as,
\begin{center}
$\frac{\mathrm{d} w_{xy}}{\mathrm{d} t} = m\frac{s_y+A}{\sum_{z \neq x}s_z+A} \times m\frac{s_x+A}{\sum_{z \neq y}s_z+A}$
\end{center}

Similar to attractiveness, the competitiveness of a node is also a very important feature. \cite{guo2011universality, jin2014co, guo2015universality} studied the role of this feature while designing competitive networks.

\subsection{Edge Preferential Attachment Model}

Chen et al. \cite{chen2007weighted} studied the phenomenon where already existing edges attract new nodes. For example in a social network, if two people are good friends, then a person who joins this network would like to be friends with them. They proposed a model by considering \textit{edge preferential attachment}, where a highly weighted edge is more powerful to attract more connection. At every time step, an edge $xy$ is selected preferentially and a new node is connected with its both endpoints. The preferential probability to select an edge $xy$ is defined as,
\begin{center}
$\prod (xy) = \frac{(1-q)w_{xy} + q}{\sum_{mn}((1-q)w_{mn}+q)}$
\end{center}
where, $q \in [0,1]$ and it helps to control the power law exponents of strength, degree and weight distributions. The weight of the selected edge $(x,y)$ is increased by $1$ and newly connected edges have weight $1$. If an edge $(x,y)$ is selected preferentially, the degree of its endpoints is increased by $1$ and the strength of its endpoints is increased by $2$. So, this model provides a linear correlation between strength and degree, given as, 
\begin{center}
$s_x=2k_x-2$.
\end{center}

Dorogovtsev et al. \cite{dorogovtsev2004minimal} presented two models, (i) generalized, and (ii) fractal model, based on this approach and their previous work \cite{dorogovtsev2000generic}. In the generalized model, the network starts with some seed graph. At every time step, one edge is selected preferentially, and its weight is increased by $w_0$. Then one node is added to the network and it is connected to both ends of the selected edge with probability $p_1$ or connected with only one end with probability $p_2$ or not added to the network with probability $p_3$, and $p_1+p_2+p_3 =1$. To select an edge preferentially, one can randomly select a node and then choose one of its edges preferentially or choose an edge preferentially among all the edges. In the fractal model, at every time step, the weight of each edge is increased according to the formula: $w=w(1+w_0)$, where $w_0$ is a constant. The triangles are made on each edge by connecting a node with both endpoints of the edge. Weights of new edges are set as, $w=1$. 
Li et al. \cite{li2013weighted} also studied the edge weight preferential attachment with the local world mechanism. The authors defined the local world containing the edges instead of nodes.

\subsection{Geographical constraint Model}
In real-world networks, such as Internet network, road network, airport network, power grid network, the spatial distance is a major constraint. In these networks, the nodes are placed in a well-defined position, and most of the edges are connected among the geographically nearby nodes. There are also some long-distance edges driven by preferential attachment to connect hubs \cite{barthelemy2003crossover}. The cost of making long-distance connections is higher than the cost of making short-distance connections \cite{xulvi2002evolving}. 
The main parameters considered while making new connections in technological or spatial networks are the strength of a node, the cost of making a new connection, and the benefits of this new connection (cost-benefit analysis). 

Barrat et al. \cite{barrat2005effects} proposed the first model to generate weighted spatial networks. The authors proposed a specific probability function to make the connections, when nodes are embedded in two dimensional space, and it is defined as,
\begin{center}
$\prod(x, y)=\frac{s_ye^{-d_{xy}/r_c}}{\sum_{z}s_ze^{-d_{xz}/r_c}}$
\end{center}
where, $d_{xy}$ is the Euclidean distance between node $x$ and $y$, and $r_c$ is the scaling factor. The network grows same as the BBV model using the proposed probability function.

Mukherjee et al. \cite{mukherjee2006weighted} proposed a weighted evolving model where nodes are connected to the nearest link. The network starts with two random points on the geographic plain connected by an edge. As the network grows, nodes are placed randomly on the plain and connected to the network. To connect a node, the nearest edge is selected, and its one of the two endpoints is connected with equal probability. The edge weight depends non-linearly on the Euclidean distance. the strength of a node can have a linear or non-linear correlation with edge weight and defined as,
\begin{center}
$s_x=\sum_{y}w^\alpha_{xy}$
\end{center}
where $\alpha$ is a constant parameter to control the impact of edge weights on the node strength. This model follows all power-law distributions observed in real-world weighted networks.

Wenhai et al. \cite{wenhai2008weighted} designed an evolving model for geographical networks. The model starts with an initial configuration of $n_0$ nodes connected by $m_0$ edges. Nodes are distributed on the plain uniformly at random. At each timestamp, a new node $x$ is added with $m$ edges and the probability of an existing node $y$ to attract new connection is given as,
\begin{center}
$\prod(x,y)= \frac{s_yf(D_{xy})}{\sum_z s_zf(D_{xz})}$
\end{center}
where, $f(D_{xy})$ is the cost function to establish a connection between node $x$ and $y$, and it depends on the Euclidean distance $D_{xy}$ of its end points, $f(D_{xy}) \propto (D_{xy})^{-\alpha}$. At each timestamp, all possible edges update their weights using the strength coupling mechanism \cite{xie2007modeling}.

Hu et al. \cite{hu2010evolution} proposed a similar model based on the interaction-force preferential attachment probability, defined as,
\begin{center}
$F(x,y) = s_xs_yd_{x,y}^{-\alpha}$.
\end{center}

In 2009, Qian and Han \cite{qian2009spatial} studied the spatial constraint on real-world weighted traffic networks. They proposed a variation of gravity equation to understand the expected flow of traffic using fitness and Euclidean distance of the nodes. It is,
\begin{center}
$T(x,y) = K\frac{M_x^{\alpha}M_y^{\alpha}}{D_{xy}^{\gamma}}$
\end{center}
where, $K$ is a constant, $M_x$ is the fitness of node $x$ and $D_{xy}$ is the Euclidean distance of node $x$ and $y$. By varying values of $\alpha$ and $\gamma$, the dependency of fitness and distance can be controlled, respectively. In real-world networks, expected traffic $T(x,y)$ can also be impacted by the traffic evolution on other nodes than its endpoints. To design the model, we assume that there are $n$ nodes uniformly embedded on the plain. Their fitness value is chosen from the given distribution $\rho(\eta)$. The Euclidean distance between two nodes can be calculated using a geometric equation. Now, the node pairs are connected one by one, which has large expected traffic. Thus the busier links are created first in the system.  The node strength depends on the traffic directly flowing from that node and the traffic passing through that node. In the traffic flow system, we always assume that the traffic flows using the shortest distance paths. So, we can calculate the traffic passing through a node by calculating the shortest paths passing through that node and the fraction of traffic that each shortest path carries. This model is helpful in understanding traffic dynamics and resource management.

\subsection{Age Weighted Model}

In unweighted networks, when a new node joins and makes connections, equal importance is given to all the links. However, in real life, all these links are not equally important. If a person is making a connection with an older node, it can be more important than a connection with a newer node. 

To explain this phenomenon, Je{\.z}ewski \cite{jezewski2004scaling} presented a model where, edge weight depends on the time interval of the birth time of both the endpoint nodes. In this model, a new node $x$ is added at every timestamp $t$ with birth time $t_x$ using degree preferential attachment. The weight $w_{xy}$ of a link $(x,y)$ depends on origin time of both end nodes, as $t_x$ and $t_y$. So, if $t_x > t_y$, edge weight $w_{xy}$ is,
\begin{center}
$w_{xy} =c_N^{-1}t_{xy}^{\sigma_x} $
\end{center}
and,
\begin{center}
$t_{xy} = t_x-t_y, $\\
$\sigma_{x} = a+b \cdot r_x,  r_x \epsilon [0,1]$\\
$c_N= c_N^{a+b\left \langle r_x \right \rangle_N},  c>0$
\end{center}
where, $N$ is the total number of nodes in the network at timestamp $t$, and parameters $a, b,$ and $c$ are independent of total nodes $N$. $\left \langle r_x \right \rangle_N$ is average of all $r_x$ so, $\left \langle r_x \right \rangle_N = (\sum_{x \epsilon N}r_x)/N$. A user can generate different scale-free power law networks using different values of weight exponent $\sigma_{x}$ and the factor $c_N$. The author further studied weighted networks and proposed an evolving model where, edge weight $w_{xy}$ depends on the degree of its endpoints $w_{xy}= (k_xk_y)^\theta$ and $\theta \in (-1,0]$ \cite{jezewski2007emergence}. This model was motivated by weak interactions of highly connected nodes where, a higher degree node pays less attention to its neighbor nodes and a lower degree node pays more attention and have strong bonds \cite{indekeu2004special, giuraniuc2005trading}. The degrees and edge weights follow power law distribution in the generated networks.



Tian et al. \cite{tian2007rank} proposed a model based on the ranking of the nodes. They defined the probability function of a node $x$ to get a new connection as, 
\begin{center}
$\prod(x) = \frac{R_x^{-\alpha}}{\sum_{y}R_y^{-\alpha}}$
\end{center}
where $R_x$ is the ranking of the node $x$. The rest of the model works as the basic BBV model except the probability function. The authors showed that there is a high correlation between the node strength and age-based ranking of the nodes. 

In 2009, Zhou et al. \cite{zhou2009age} proposed an age based model using mutual selection approach. The model starts with $n_0$ isolated nodes with initial age $h=1$. At each timestamp, a new node is added and each existing node selects $m$ other nodes. Node $x$ selects node $y$ with the probability,
\begin{center}
$\prod(x,y) = \frac{h_y^{-\alpha}}{\sum_{z}h_z^{-\alpha}-h_x^{-\alpha}}$
\end{center}
where, $\alpha>0$ and controls the effect of node strength in getting new connections. By varying the value of $\alpha$, one can generate assortative or disassortative networks. Growth is based on mutual selection, so, connections are not updated, unless two nodes select each other mutually. In these networks, a younger node has a higher age coefficient $h$ and has a low probability to get more connections. 

Wen et al. \cite{wen2011weighted} combined the aging phenomenon with the concept of the local world. In this model, the local world is determined as in \cite{pan2006generalized}. At each time step, $M$ nodes are selected randomly to constitute the local world of the new node. Then a new node $x$ is added to the network with $m$ edges, and $m$ is chosen randomly from the given range $m \epsilon [1,q]$. Each new age is connected to a node $y$ with strength age preferential attachment probability,
\begin{center}
$\prod(x,y) = \frac{s(t_x,t) \cdot (t-t_x^{-\alpha})}{\sum_{z \epsilon M}s(t_z,t) \cdot (t-t_z^{-\alpha})}$
\end{center}
where, $t_x$ is the birth time of node $x$ and $s(t_x,t)$ denotes the strength of node $x$ at time $t$. $\alpha$ is a decay factor to control the effect of aging, and $\alpha \epsilon [0,1)$. Weights are rearranged after adding each node. This model can be efficiently used to reproduce Internet or citation networks.



\section{Signed Weighted Networks}


In real-world networks, all relationships can not be captured by positive edge weights. In some networks, such as frenemy networks or trust-distrust networks, there are two kinds of edges, positive and negative, that denote the positivity or negativity of the relationship, respectively. For example, in real-world social networks, we have a positive or negative opinion about a person based on the bonding or the first impression. Researchers have studied such networks to understand their evolution and the evolution of positive and negative edge-weights. An analysis of these networks shows whether the environment is hostile or harmonious. For an organization or institution, a harmonious environment is essential and desired. Researchers have proposed several models to generate synthetic signed networks that we will discuss next. 

In 2004, Hu et al. \cite{hu2004model, hu2005weighted} proposed a model to understand students' relationship in a class \cite{saxena2019survey}. The model starts with $N$ nodes, where everyone knows every other person. The strength of the relationships is stored in $n \times n$ adjacency matrix. Positive edge weights show harmonious relations, and negative edge weights show hostile relations. The model tries to capture the impact of several encounters between two people. In the modeling, the value $1$ is assigned with probability $p$ and $-1$ with probability $(1-p)$ to the elements of the matrix. 
At every time step, the following steps are executed: 
\begin{enumerate}
\item A node $x$ is selected randomly and $x$ chooses one other node $y$ for the interaction using the following given probability function:
\begin{center}
$P(xy) = \frac{|w_{xy}| +1}{\sum_{y}(|w_{xy}|+1)}$
\end{center}
\item After choosing $x$ and $y$, $w_{xy}$ is changed by $\pm1$ with some probability. For getting the value of this probability, a parameter $\gamma_{xy}$ is defined based on the similarity of their relationships. If two people have more number of mutual friends or enemies, it is more likely that they will become good friends. 
\begin{center}
$\gamma_{xy} = C^{-1}\sum_{z}w_{xz}\cdot w_{zy}$
\end{center}
where,
\begin{center}
$C=\sqrt{\sum_{z}w_{xz}^2}\cdot \sqrt{\sum_{z'}w_{yz'}^2}$
\end{center}
\end{enumerate}
So, $\gamma_{xy}  \in  [-1,+1]$ . If $\gamma_{xy} > 0$, the value of edge weight is increased by $1$ with probability $\gamma_{xy}$ and if $\gamma_{xy} < 0$, the value of edge weight is decreased by $1$ with probability $|\gamma_{xy}|$. 

The harmoniousness of the whole network is computed using,
\begin{center}
$\sum_{x,y,w_{xy}>0}\frac{w_{xy}}{\sum_{x,y}|w_{xy}|}$
\end{center}
Similarly, the hostility of the network can be calculated. The authors also observed the critical value of $p$, and above this value, the environment starts to converge to more harmony. They further showed that this model follows real-world properties.

\section{Mesoscale Structured Weighted Networks}

Real-world networks have mesoscale structural properties, such as community and core-periphery structure. Next, we will discuss evolving models to generate mesoscale structured weighted networks. 

\subsection{Community Structured Weighted Networks}

A community is a group of nodes having dense connections within and sparse connections outside. In community structured networks, links can be categorized as, (i) \textit{intra-community link} if both end-nodes belong to the same community, otherwise (ii) \textit{inter-community link}. Newman \cite{newman2004analysis} studied the community structure in weighted networks and modified the modularity function to identify communities in weighted networks. Lancichinetti and Fortunato \cite{lancichinetti2009benchmarks} proposed a benchmark to design directed and weighted networks with defined community structure. The different model parameters, such as the number of communities, the number of nodes in each community, and get the community structured graph. Some models have been proposed to generate weighted networks with the community structure that we will discuss next. 

Li and Chen \cite{li2006modelling} proposed the first model with inbuilt community structure based on inner-community and inter-community preferential attachment. Let's assume, $M$ is the total number of community in the network. The preferential attachment rule is defined as,
\begin{enumerate}
\item \textbf{Inner-Community Preferential Attachment:} When a node is chosen in a randomly selected community, we only consider inner-community degree for the preferential attachment. $k^1_{xi}$ is the inner degree of a node $x$ in community $i$. So, node $x$ will be chosen with probability,
\begin{center}
$\prod (k^1_{xi}) = \frac{k^1_{xi}}{\sum_{y}k^1_{yi}}$
\end{center}
\item \textbf{Inter-Community Preferential Attachment:} When a new node is making connections with nodes of other communities, it only considers inter community degree of that node. $k^2_{xj}$ is the inter degree of a node $x$ in community $j$. A node $x$ in community $j$ will be chosen with the probability,
\begin{center}
$\prod (k^2_{xj}) = \frac{k^2_{xj}}{\sum_{m,n,n \neq i}k^2_{m,n}}$
\end{center}
\end{enumerate}
Similarly, the preferential strengthening rules are defined as,
\begin{enumerate}
\item \textbf{Inner-Community Strengthening:} The probability to choose nodes $x$ and $y$ from community $i$ is defined as,
\begin{center}
$\prod (x,y,i) = \frac{w^1_{xyi}}{\sum_{m,n}w^1_{mni}}$
\end{center}
\item \textbf{Inter-Community Strengthening:} The probability to choose two existing node $x$ and $y$ belonging to two different communities $i$ and $j$, respectively, is defined as,
\begin{center}
$\prod (x,i,y,j) = \frac{w^2_{xiyj}}{\sum_{m,p,n,q}w^2_{mpnq}}$
\end{center}
\end{enumerate}

The model works in the following manner,
\begin{enumerate}
\item \textbf{Initialization:} the network starts with $M$ seed communities, each having $m_0$ fully connected nodes. These communities are connected with each other using $M(M-1)/2$ inter-community links, where each community is connected to $(M-1)$ communities.
\item \textbf{Growth:}
\begin{enumerate}
\item With some probability $\alpha$, a community is selected uniformly at random, and a new node is added to it using the inner-community preferential attachment rule. With some probability $\beta$, this node will also make some inter-community links. All added links have weight $1$. 
\item  With probability $1 - \alpha$, no new node will be added. A new inner-community link is added. With probability $\eta$, a new inter-community link is added. Both links are added with weight $1$ and using preferential strengthening rule.
\end{enumerate}
\end{enumerate}
The value of $\beta$ and $\eta$ is small, so that inter-community connections are sparse. In this model, inner-degree, inter-degree, total degree, link weights, and strength follow a power-law distribution.

In 2007,Kumpula \cite{kumpula2007emergence} proposed a model based on the following two phenomenon, 
\begin{enumerate}
    \item \textbf{Cyclic Closure:} Friends of friends \cite{hanaki2007cooperation}
    \item \textbf{Focal Closure:} Make friends randomly through some consequences \cite{kossinets2006empirical}.
\end{enumerate}
The model starts with a fixed size network of $N$ nodes, and the network grows by the following given steps.
\begin{enumerate}
\item \textbf{Local Attachment:} Local attachment follows the concept of triadic closure. In time interval $\Delta t$, each node $x$ having at least one neighbor starts a weighted local search and reaches to $y$. If node $y$ has any other neighbor than $x$, $y$ starts a weighted local search and stops at $z$. If there is no link between $x$ and $z$, link $(x,z)$ is formed having weight $w_0$ with some probability. If the link exists, its weight is increased by $\delta$. In both cases, link weight for $w_{xy}$ and $w_{yz}$ is also increased by $\delta$.
\item \textbf{Global Attachment:} A node having no links creates a link with a random node selected using the given probability, and the edge weight is $w_0$.
\item \textbf{Node deletion:} In this step, a node is deleted with some probability and replaced with a new node having no links, so that system size is maintained.
\end{enumerate}
The value of $\delta$ should be large to get tightly knit communities. The networks generated using this model follow all the properties of complex weighted networks and are assortative. They extended their work \cite{kumpula2009model} and showed that this algorithm could be implemented very efficiently and CPU-time is linearly dependent on network size. Jo study the community structure and bursty dynamics evolving phenomenon using the local attachment, global attachment, and human dynamics \cite{jo2011emergence}. This model shows a heavy tailed inter-event time distribution.


In real-world social networks, relationships are faded or broke up with time. To implement these phenomenon, \cite{murase2015modeling} proposed a model by extending the work of Kumpula \cite{kumpula2007emergence}. This model modified the node deletion step by using two different approaches, (i) link deletion (to capture the broke up of relationships) and (ii) link aging (to capture the gradual fading of relationship). For the link deletion, at each time step, all links are removed with some probability to show the sudden break-up. At each time step, the weight of all the links is decreased by a constant factor to implement the link aging. If link weight is less than the threshold value, the link is removed from the system. 

Zhou et al. \cite{zhou2008weighted} presented a model where the community size also follows power-law distribution as observed in real-world networks \cite{clauset2004finding}.  
The model works on the following steps.
\begin{enumerate}
\item The network is initialized as in \cite{li2006modelling} model.
\item \textbf{Growth Phase:} In the growing phase, a new node or a new community will be added as described in steps (a) and (b).
\begin{enumerate}
\item A new node is added to the network with probability $p$, and it makes $m$ connections in the network.  When a new node is added to the network, it makes following two types of connections.
\begin{enumerate}
\item \textbf{PA Links:} A new node $x$ first chooses a community $i$ using community size preferential attachment (bigger the size of the community higher the probability to be selected). The community size preferential attachment probability is defined as,
\begin{center}
$\prod (S_i) = \frac{S_i}{\sum_{j}S_j}$
\end{center}
where $S_i$ is the size of community $i$. Then, it chooses a node $y$ in community $i$ using PA rule. This selected community creates the local world of the new coming node.
\item \textbf{TF Links:} 
The node $x$ makes triad formation links with neighbors of $y$ using PA. The node $x$ makes one PA connection and remaining $(m-1)$ are PA links with probability $\varphi$ or TF links with probability $(1-\varphi)$. PA links are formed as inner-community link with probability $q$ and inter-community link with probability $(1-q)$. In both steps (i and ii), all links are added with weight $w_0=1$.
\end{enumerate}
\item A new community is added to the network with probability $(1-p)$. One node is chosen randomly from this new community and makes $m$ connections with other communities. The links are added in the same way as in the above steps, except no inner-community link will be added.
\end{enumerate}
\item \textbf{Weight Rearrangement:} Each new link $(x,y)$ creates extra traffic $\delta$, and it is proportionally distributed among all outgoing edges of node $y$.
\end{enumerate}

\subsection{Core-periphery Structured Weighted Networks}

In a network, the nodes follow a hierarchical order that gives rise to the core-periphery structure. The core is a dense central nucleus of the network. For example, in a co-authorship network, the core nodes are pioneer researchers of the area. The nodes belonging to the core are highly connected and highly connected with periphery nodes. 

Saxena and Iyengar \cite{saxena2016evolving} studied the evolution of core in social networks and observed that periphery nodes make a high number of connections with other core nodes to move into the core of the network. They further proposed an evolving model to generate weighted networks having both community as well as core-periphery structure based on their observations. In the proposed model, the network evolves using the preferential attachment and triad formation method. The network starts with a seed graph having $c$ communities, and each community has a clique of $n_0$ number of nodes. All nodes of the seed graph are considered core nodes. At each timestamp, the following steps are executed.
\begin{enumerate}
    \item A community is selected uniformly at random, and a new periphery node $x$ is added to this community. The node $X$ makes $m \cdot f$ intra-community connections using intra-community strength preferential attachment, and $m \cdot(1-f)$ inter-community connections using inter-community strength preferential attachment, where $m$ is the total number of connections created by $x$ and $f$ denotes the fraction of intra-community connections.. When nodes $x$ and $y$ are connected, the BBV weight distribution function is used to balance the weights.
    \item With probability $q$, a periphery node $y$ is selected using strength preferential attachment rule to move to the core of the network. The node $y$ is connected with each core node with probability $p$.
    \item Each existing node in the network makes triad formation connections with probability $r$ using strength preferential attachment rule. If the selected node is already connected with the given node, then their edge weight is increased by $w_0$.
\end{enumerate}

Each new connection is initialized with weight $w_0 = 1$. In the proposed model, step 1 regulates the topological growth, and steps 2 and 3 regulate the self-growth of the network. Specifically, step 2 manages the evolution of the global core. The probability $p$ controls the density of the core, and so it has a high value. The probability $q$ controls the size of the core, and the probability $r$ controls the overall self-growth and has a small value. The authors showed that the generated networks have properties similar to those observed in real-world networks. The other models to generate mesoscale structured networks are \cite{saxena2015understanding,  gupta2016modeling, gupta2019modeling}.

\section{Multilayered Weighted Networks}

Multilayered networks are used to represent an environment where each node interacts with others in diverse contexts. Each layer represents a single interacting context, such as friendship, colleagues, family, etc. 
The nodes in different layers are the same, and the relationship in one layer is corresponding to one context. A multilayered network can be represented as $G= (G_1, G_2, G_3, ... G_k)$, where each graph $G_i$ $(1 \leq i \leq n)$ represents a single-layered network and can be stored using weighted adjacency matrix.

Murase et al. \cite{murase2014multilayer} proposed a model to design multilayered weighted networks while maintaining the spatial constraint and Granovetter structure. Each layer of the network is designed using the single-layer generative model proposed by Kumpula et al. \cite{kumpula2009model}. The global attachment follows the spatial preferential attachment, and two nearby nodes have a high probability of being connected. The authors used the percolation method to study these networks' properties and observed that geographical proximity plays an important role in designing such networks. In \cite{menichetti2014correlations}, authors studied the correlation between network topology and edge weights in multilayered networks and proposed a framework to design null models of multiplex networks. 

\section{Application Specific Weighted Models}

All the above-discussed models are generalized models and can be used to create different types of weighted networks. Researchers also studied the application-specific weighted networks and proposed modeling approaches for these networks. In this section, we will explain various application-specific models.

\subsection{International Trade Network}

Bhattacharya et al. \cite{bhattacharya2008international} observed properties of \textit{International Trade Networks (ITN)}. In these networks, few rich countries (called rich club) control the major fraction of the global trade. The size of the rich club is shrinking with time. In the ITN, nodes are the countries, and edge weight shows the trade volume between the two countries. Trade volume depends on the export and import volumes of both the countries. They proposed a model based on the gravity law. The gravity equation is defined as,
\begin{center}
$F_{xy}= m_x^\alpha(\frac{m_y^\beta}{l_{xy}^\theta}/\sum_{z \neq x}\frac{m_z^\beta}{l_{xz}^\theta})$
\end{center}
where, $m_x$ denotes the economic size and $l_{xy}$ denotes the distance of two economic centres. Model starts with $n$ points uniformly distributed on the unit square plain. Each node is assigned a random gross domestic product (GDP) such that the total GDP is $1$.  At each time step, a pair of countries is selected and its edge weight is updated using total trade transaction. After each transaction update, all nodes update their GDP to maintain the total GDP as unity. Using this model, real-world ITN networks can be designed. 

\subsection{Bipartite Weighted Trade Model}

Chakraborty and Manna \cite{chakraborty2009weighted} proposed a \textit{bipartite weighted trade model} with quenched random saving propensities. This model is inspired by their previous work of the kinetic exchange model \cite{chatterjee2007kinetic}. The model starts with the fixed number of traders, each having a fixed assign wealth $w$. At each time step, A pair of traders $x$ and $y$ is selected with the probability defined as,
\begin{center}
$\prod(x) = w_x^\alpha$ and $\prod(y) = w_y^\beta$
\end{center}
where, $\alpha, \beta \geq 0$. After selecting a pair, trading happens between them, and wealth values are updated. They also showed that wealth distribution follows Pareto law.

\cite{riccaboni2014stochastic} proposed a stochastic model to fit International Trade flow better. This model is based on two main approaches, (i) the first controls the formation of trade links, and (ii) the second controls the trade intensity. Other models of ITN are proposed by \cite{duenas2013modeling, almog2015gdp}.

\subsection{Rail Transportation Network}

Wan et al. \cite{wan2014weighted} proposed a traffic driven weighted evolving model for \textit{rail transportation network} based on the city attractiveness. City attractiveness depends on the urban passenger traffic flow and the fitness of the city. The network starts with a seed graph. At each time step, a new node $x$ is added to the network with $m$ edges. The probability of connecting an existing node $y$ is defined as, 
\begin{center}
$\prod(y)= \frac{s_y+\frac{\alpha_y}{L_y}}{\sum_{z}(s_z+\frac{\alpha_z}{L_z})}$
\end{center}
where $\alpha_y$ is the attractiveness of city $y$, and $L_y$ is the distance of the new node from the node $y$. So, the probability depends on both attractiveness of the node and the position of the node.

\subsection{Immune System Model}

Ruskin and Burns \cite{ruskin2006weighted} proposed a directed weighted bipartite model of \textit{immune system development}. Nodes are divided into two classes, (i) Immune cells (ctl) and (ii) Antigen-presenting cells (apc). An edge represents the interaction between cells. Each cell has a fixed location in shape space and can attack within the area of radius $r$. When a new node joins, the probability of connecting with an existing node depends on both its strength and location. The disparity of a node can be calculated as,
\begin{center}
$disp(x)= \sum_{y \epsilon \Gamma(x)} (w_{xy}/s_x)^2$
\end{center}
It shows the distribution of strength across edges connected to the node.

\subsection{Stock Market Network}

Qi et al. \cite{qi2013study} studied the stock market of information services share in Shanghai and Shenzhen A-share and proposed a model to reproduce directed weighted stock networks using the fitness value of the company. The network starts with a seed graph, and the evolution mechanism follows the given steps.
\begin{enumerate}
\item At each time step, a new node is added with a fixed fitness value $\mu$, drawn from a given fitness distribution $\rho(\mu)$. 
\item The new node is connected to $m$ existing nodes using given probability function,
\begin{center}
$\prod(x)= \frac{\mu_xs_x}{\sum_{z}\mu_zs_z}$
\end{center}
\item With probability $p$, weights are assigned to new edges by following power law distribution otherwise assign weight $w_{xy}=0$.
\item $am$ edges are added among already existing nodes and $(a \geq 0)$.
\item $bm$ edges are deleted randomly from the existing edges and $(b \geq 0)$.
\item An existing node is deleted and all its edges with probability $c$.
\end{enumerate}
The authors analyzed different stock networks that are generated using different fitness distribution.

\subsection{Email Network}

Valverde and Sol{\'e} \cite{valverde2006evolving} studied the \textit{Email network} of an open source software development community. They found that the communication network is highly symmetric $w_{xy} \approx w_{yx}$. The network starts with a seed graph and at every time step, a new node joins and sends $m$ emails. The probability that an old node $x$ replies to new node depends on the betweenness centrality (load) of the node ($b_x(t)$) at time $t$, given as,
\begin{center}
$\prod(x)= \frac{(b_x(t)+c)^\alpha}{\sum_{y}(b_y(t)+c)^\alpha}$
\end{center}
where, $c$ and $\alpha$ are constant and their values depend on the application. The betweenness centrality of all nodes is recalculated after each iteration.

There are some more works to generate application specific network, such as for agri-food supply chain network \cite{li2014evolving}, wireless sensor network \cite{zhang2012local,jiangstrength}, port network \cite{xiong2008study}, Internet network \cite{xuelian2011research,peng2006router}, transportation network \cite{gao2007asymmetry, hai2015research, li2006weighted, earnest2011geographic}, telephone call network \cite{tam2007modeling}, urban growth model \cite{rui2013urban}, energy supply-demand network \cite{sun2013general}, collaboration network \cite{li2007econophysicists}, International migration model \cite{letouze2009revisiting, fagiolo2013international}, crime location network \cite{qian2011weighted1}, earthquake network \cite{zhang2009self}, power grid network \cite{changhong2009evaluation, shen2012weighted}, and gene co-expression network \cite{wang2012research}.

\section{Other Miscellaneous Models}


Here, we cover models that can not be categorized in the above categories.

\subsection{Random Network Model}

Garlaschelli \cite{garlaschelli2009weighted} proposed a weighted random graph model based on Erdos- Renyi model \cite{erd6s1960evolution}, where, the probability to put an edge of weight $w$ between node $x$ and $y$ is defined as,
\begin{center}
$P_{xy}(w)=(\gamma_x\gamma_y)^w(1-\gamma_x\gamma_y)$
\end{center}
where, $\gamma_x < 1$ and it controls the expected strength of node $x$.

Meng and Zhu \cite{meng2011approach} also presented a designing approach based on the Erdos-Renyi model. To design a network, first create an Erdos-Renyi graph where all edges carry equal weight. At each iteration, edge weights are redistributed such that the total weight of all edges is constant. Once the network is stabilized, the degree and strength follow a power-law distribution. 

In most of the discussed models, a new node is attached preferentially, and the self-growth is also governed by the preferential law. Zhang et al. \cite{zhang2012model} proposed a model where a new coming node makes links randomly. When a link is connected randomly to a node $x$, its strength is increased by $ps_x$, where $(p>0)$ and $p$ is the innovative ability of nodes. This increased strength is distributed among all the connected links preferentially, as $w_{xy}(t+1) = w_{xy}(t) + pw_{xy}(t)$. As a new node is connected randomly, degree distribution does not follow the power law, but as high strength nodes get more strength due to the multiplicative nature of strength increase, strength and link weights follow a power-law distribution.

\subsection{Random Weight Assignment Model}

Park et al. \cite{park2004characterization} proposed a very simple and minimal \textit{random weight assignment model} as an extension of the BA model to design weighted networks. The model is based on the concept of node betweenness and the scaling of betweenness with nodes' strength. The model first generates a scale-free unweighted network using the Barabasi-Albert model \cite{barabasi1999emergence} of $n$ nodes. To convert it to the weighted network, first, we assign the number $0-n$ to each node randomly and then divide this value by $n$ to normalize all values in the preferred range $[0,1]$. Now, the weight of a link $w_{xy}$ is computed as $(w_x+w_y)/2$. This edge weight value shows the traffic transferred through the given edge when traffic flows between every two nodes using optimal path \cite{braunstein2003optimal}. To improve the quality of this model, the nodes' weight can be assigned using the intrinsic properties of the nodes.



\subsection{Weighted Stochastic Block Model}

Aicher et al. \cite{aicher2013adapting} studied \textit{weighted stochastic block model} (WSBM) where edge weights are picked from an exponential family distribution. This model depends on two parameters $z$ and $p$, where $z$ is a vector of size $n$ containing block assignment of each node and $p$ is a $n \times n$ matrix containing the probabilities $p_{xy}$ of connecting two nodes $x$ and $y$. Different values of $p$ will generate assortative, disassortative, multi-partite, hierarchical, or core-periphery structure. 

\subsection{Weighted Fractal Networks}

Carletti and Righi \cite{carletti2010weighted} presented the first detailed evolving model for \textit{weighted fractal networks} (WFN). These networks are characterized by two main parameters $s$ and $f$, where, $s$ $(s>0)$ is the number of copies, and $f$ $(0<f<1)$ is the scaling factor. The network starts with an initial weighted graph $G_0$ and a family of WFN is constructed using predefined mapping $G_k = T_{s,f,a}(G_{k-1})$. $T_{s,f,a}$ is a mapping function based on the scaling factor, the number of copies, and a labelled node $a$ that is also called attaching node. 
In WFN, the node strength has a power-law probability distribution where the exponent is the Hausdorff dimension \cite{mandelbrot1983fractal}. They further studied non-homogeneous WFN, where different scaling is defined for all edges. They extended this work and proposed deterministic non-homogeneous and stochastic weighted fractal network models \cite{carletti2010stochastic}. 

Zhang et al. \cite{zhang2010deterministic} constructed a deterministic fractal network and investigated that its recursive structure helps in the exponential decay of traffic fluctuations of nodes and edges. To further understand the fractal networks and their dynamics \cite{dong2015non} can be referred.

\subsection{Capacity Constraint Model}

Wu et al. \cite{wu2013weighted} presented a model based on the capacity constraint of the nodes. A node can not be connected with more than its capacity. For each node its attractive index $A_x(t)$ at time $t$ is defined using its capacity $C_x$, as,
\begin{center}
$A_x(t)=s_x(t)(1-\frac{s_x(t)}{C_x})$
\end{center}
At each timestamp, a new node is added and make $m$ connections using the attractive index preferential attachment. The increased weight is distributed proportionally to the attractiveness of the other endpoint.

\subsection{Recursive Model for Hierarchical Weighted Networks}

Sun et al. \cite{sun2014scaling} proposed a recursive model to generate hierarchical weighted network $H_k$ containing $n^k$ vertices. Each vertex is represented by a $k$ length tuple as $x_1x_2x_3......x_k$, where, each $x_i \epsilon S$ and $S=\left \{0, 1, 2, ... , n-1\right \}$. the model is defined as:
\begin{enumerate}
\item $H_1$ is the initial graph having $n$ nodes and $(n-1)$ edges of weight $1$.
\item $H_k$ can be obtained by merging $(n-1)$ copies of  $H_{k-1}$whose edge weights are updated by a constant factor and original $H_{k-1}$. New edges are added by following below given steps:
\begin{enumerate}
\item Put an edge $(x_1x_2x_3...x_{k-1}x_k, x_1x_2x_3...x_{k-1}x_k)$, where $x_k \neq y_k$ and $x_k=0$, $y_k \neq 0$.
\item Put an edge $(x_1x_2x_3...x_i00...0, x_1x_2x_3...x_ix_{i+1}...x_k)$, where $x_j \neq 0$, $i+1 \leq j \leq k$, $1 \leq i \leq k-2$.
\end{enumerate}
\end{enumerate}
The authors also showed that the average weighted shortest path (AWST) and the average receiving time (ART) stay bounded with the network size.

\cite{dai2013random} also proposed a model for non-homogeneous weighted Koch networks and studied AWST and AWRT on these networks. A family of networks can be generated by applying mapping on the given initial network. The initial graph $G_0$ consists of three nodes connected by three edges of unit weight (triangular structure). $G_t$ can be generated from $G_{t-1}$ such that it contains $4^t$ triangles, $2 \times 4^t+1$ nodes, and $3 \times 4^t$ edges.

\subsection{Overlapped Clique Evolution Model}

Yang et al. \cite{yang2010novel} proposed a \textit{weighted clique evolution model} for overlapped clique growth. This model starts with $c_0$ cliques, each having $n_0$ vertices and all the edges are having weight $w_0=1$. At each time step, a new clique $c$ is added with $n$ nodes. A new clique contains $n_1$ old nodes and $n_2$ new nodes; old nodes are selected using the following selection mechanism:
\begin{enumerate}
\item \textbf{Preferential selection mechanism:} Probability to select a vertex $x$ depends on the number of cliques $h_x$ that the vertex joins. It is defined as, 
\begin{center}
$\prod(x) = \frac{h_x}{\sum_z h_z}$
\end{center}
\item \textbf{Random selection mechanism:} An old vertex is selected randomly from all the existing vertices.
\end{enumerate}
The distribution of the number of cliques that a vertex joins $P(h)$ follows a power law. They also studied three bus transportation networks of Shanghai, Hangzhou, and Beijing, China, and showed that they are similar to overlapping clique networks generated using the proposed model. \cite{xu2011modeling, li2013dynamically} also studied similar weighted clique evolution models. 


\section{Conclusion}

In this chapter, we have discussed the evolution of different kinds of weighted networks and evolving models to explain the underlying evolutionary mechanisms. The chapter presented a systematic review of different modeling frameworks. We further discussed the modeling phenomenon for inbuilt mesoscale properties, such as community structure or hierarchical organization. We also covered different modeling approaches that are proposed to generate real-world networks having special properties. The chapter will help better understand the structural properties of real-world networks and dynamic phenomenon taking place on these weighted networks. 

\bibliographystyle{unsrt}
\bibliography{mybib}

\end{document}